\documentclass[aps,prl,showpacs,superscriptaddress,groupedaddress,preprint]{revtex4-1} 
\usepackage{amsmath}    
\usepackage{amssymb}
\usepackage{graphicx}   
\usepackage{verbatim}   
\usepackage{color}      
\usepackage{subfigure}  
\usepackage[colorlinks=true,linkcolor=blue,citecolor=blue,urlcolor=blue]{hyperref}   
\usepackage{epstopdf}
\usepackage{float}
\usepackage{xcolor}
\definecolor{YellowOrange}{HTML}{FFA500}

\begin{document}

\title{Energy invariance in capillary systems}

\author{\'Elfego Ruiz-Guti\'errez}
\affiliation{Smart Materials \& Surfaces Laboratory, Faculty of Engineering and Environment, Northumbria University, Ellison Place, Newcastle upon Tyne, NE1 8ST, United Kingdom}
\author{Jian H. Guan}
\affiliation{Smart Materials \& Surfaces Laboratory, Faculty of Engineering and Environment, Northumbria University, Ellison Place, Newcastle upon Tyne, NE1 8ST, United Kingdom}
\author{Ben Xu}
\affiliation{Smart Materials \& Surfaces Laboratory, Faculty of Engineering and Environment, Northumbria University, Ellison Place, Newcastle upon Tyne, NE1 8ST, United Kingdom}
\author{Glen McHale}
\affiliation{Smart Materials \& Surfaces Laboratory, Faculty of Engineering and Environment, Northumbria University, Ellison Place, Newcastle upon Tyne, NE1 8ST, United Kingdom}
\author{Gary G. Wells}
\affiliation{Smart Materials \& Surfaces Laboratory, Faculty of Engineering and Environment, Northumbria University, Ellison Place, Newcastle upon Tyne, NE1 8ST, United Kingdom}
\author{Rodrigo Ledesma-Aguilar}
\affiliation{Smart Materials \& Surfaces Laboratory, Faculty of Engineering and Environment, Northumbria University, Ellison Place, Newcastle upon Tyne, NE1 8ST, United Kingdom}

\date{\today}

\begin{abstract}

We demonstrate the continuous translational invariance of the energy of a capillary surface in contact with reconfigurable solid boundaries.  
We present a theoretical approach to find the energy-invariant equilibria of spherical capillary surfaces in contact with solid boundaries of arbitrary shape and examine the implications of dynamic frictional forces upon a reconfiguration of the boundaries. 
Experimentally, we realise our ideas by manipulating the position of a droplet in a wedge geometry using lubricant-impregnated solid surfaces, which eliminate the contact-angle hysteresis and provide a test bed for quantifying dissipative losses out of equilibrium. 
Our experiments show that dissipative energy losses for an otherwise energy-invariant reconfiguration are relatively small, provided that the actuation timescale is longer than the typical relaxation timescale of the capillary surface.   
We discuss the wider applicability of our ideas as a pathway for liquid manipulation at no potential energy cost in low-pinning, low-friction situations. 
\end{abstract}

\maketitle

{\it Introduction.--}
Capillary surfaces, which are infinitely thin surfaces that separate two fluids, are an everyday example of how beautiful symmetrical shapes appear in nature.
Because they store surface energy, capillary surfaces underpin important physical phenomena
such as the extreme superhydrophobicity exhibited by many plant and animal species, 
the internal adhesion of granular media, 
and the stability of foams and emulsions~\cite{bonn2009wetting}.

The fundamental equilibrium principle of the modern theory of capillarity is the minimisation of the total surface energy~\cite{degennes2013capillarity},
which for a solid-liquid-gas system reads $F= \gamma_{\rm lg} A_{\rm lg} + \gamma_{\rm sl} A_{\rm sl} + \gamma_{\rm sg} A_{\rm sg}$, 
where $\gamma_i$ and $A_i$ refer to the surface energy and surface area of the liquid-gas ($i={\rm lg}$), solid-liquid ($i={\rm sl}$) and solid-gas ($i={\rm sg}$) interfaces. 

As first noted by Gauss, the minimisation of $F$ is a variational problem which yields two central equations for the shape of a capillary surface. 
First, the liquid-gas interface must satisfy the Young-Laplace equation, 
\begin{equation}
\Delta p = 2 \gamma_{\rm lg} \kappa, 
\label{eq:YL}
\end{equation}
which relates the pressure difference between the liquid and the gas, $\Delta p$, to the Laplace pressure, $2\gamma_{\rm lg} \kappa$, where $\kappa$ is the mean local curvature of the interface.
Second, upon contact with a solid boundary, the interface profile must satisfy Young's Law, 
\begin{equation}
\cos \theta_{\rm e} = \frac{\gamma_{\rm sg} - \gamma_{\rm sl}}{\gamma_{\rm lg}}, 
\label{eq:Y}
\end{equation}
which determines the intersection angle with the solid, $\theta_{\rm e}$, also known as the equilibrium contact angle.

Finding solutions of the Young-Laplace equation, subject to the boundary condition imposed by Young's Law, is a paradigm in capillarity~\cite{finn2012equilibrium,Anderson2006}. 
Once an equilibrium solution is found, its stability can be examined, and the surrounding energy landscape constructed. 
A displacement from equilibrium can be static or dynamic, but, in most cases, will lead to a change in the surface energy. 
Motion can only occur if this change surpasses the static energy barrier of contact-angle hysteresis~\cite{quere2002rough}, 
and the timescale of the motion that follows is typically set by competing capillary, dissipative and external forces~\cite{bruus2008theoretical}. 

The relation between symmetry and energy invariance is a central concept across physics. 
In general, an equilibrium state with a high degree of symmetry will have a large number of energy-degenerate configurations mapping onto that state. 
For capillary surfaces, this implies the existence of energy landscapes where energy-invariant equilibria are either continuously or discreetly distributed in the parameter space 
(a familiar example is the translational symmetry of a droplet on a flat solid surface). 

It is natural to consider whether the intrinsic symmetries of a capillary surface can ensure the invariance of the surface energy upon a reconfiguration of the boundaries.
If so, one can further ask if energy-invariant trajectories that achieve a net translational motion of the capillary surface can be devised. 
This is an interesting problem from the point of view of theory, 
and, is experimentally challenging because of the barriers imposed by contact-angle hysteresis and dynamic frictional forces. 

In this Letter we demonstrate the energy-invariant translational motion of a capillary surface upon actuation of bounding solid surfaces.  
We focus on spherical surfaces as a model system, 
which  appear in numerous situations of fundamental and practical relevance. 
We first introduce a theoretical approach to find paths of energy-invariant equilibria, 
and then examine the implications of dynamic frictional forces using a Lagrangian approach. 
Experimentally, we exemplify our ideas by manipulating the position of a droplet in a wedge geometry using Slippery Liquid-Infused Porous Surfaces (SLIPS)~\cite{wong2011bioinspired}, 
also known as Lubricant-Impregnated Surfaces (LIS)~\cite{smith2013droplet}, 
which eliminate the contact-angle hysteresis and provide a test bed for quantifying dissipative losses out of equilibrium. 
We discuss the wider applicability of our results as a pathway for liquid manipulation at no surface-energy cost in low-pinning, low-friction situations. 

\begin{figure}
\includegraphics[width=0.5\columnwidth]{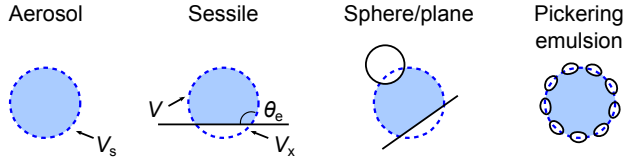}
\caption{\label{fig:Truncated} (Color online) Construction of energy-invariant equilibria of spherical capillary surfaces.}
\end{figure}

{\it Theory.--}
We start by considering the simplest solution of the Young-Laplace equation for a solid-liquid-gas system that preserves a spherical symmetry. 
This corresponds to an aerosol droplet of volume $V_{\rm s}$ and surface area $A_{\rm s}$ [see Fig.~\ref{fig:Truncated}]. 
One can map this geometry to a sessile droplet by considering the intersection of the sphere with a solid plane of total surface area $A$. 
The required droplet shape is enforced by choosing the level of the truncation, which determines the intersection angle with the solid plane, $\theta_{\rm e}$. 
This fixes the excluded volume of the sphere, $V_{\rm x}$, and the volume of the droplet, $V = V_{\rm s} - V_{\rm x}$. 
As a result, one finds the total surface energy, $F_{\rm e} = \gamma_{\rm lg} A_{\rm s} - F_{\rm x}$, 
where $F_{\rm x} = \gamma_{\rm lg} A_{\rm s}[(1+ \cos\theta_{\rm e})/2+\sin^2\theta_{\rm e}\cos\theta_{\rm e}/4]-\gamma_{\rm sg} A,$
is the free energy of the excluded cap, the solid-liquid footprint and the dry portion of the solid plane.

This construction can be generalised to include an arbitrary number of  non-intersecting boundaries, and of any shape. 
Mechanical equilibrium is guaranteed because the truncated sphere shape satisfies the Young-Laplace equation, whilst the intersection with the solid boundaries now requires that Young's Law is satisfied locally along each contact line.
With these considerations in mind, one can immediately find the static surface energy of the capillary surface, 
\begin{equation}
F_{\rm e} = \gamma_{\rm lg} A_{\rm s} - \sum F_{{\rm x} i}, 
\label{eq:F}
\end{equation}
where the first term is the contribution of the full sphere and the second term is the energy arising from the portions excluded by the solid boundaries. 

Here we shall focus on the  situation where the solid surfaces have uniform wettability~\footnote{Although we note that the more general case of heterogenous surfaces can be of significant interest, both mathematically and physically}. 
In such a case, the requirement of a constant equilibrium contact angle over a solid surface imposes the constraint that, close to the contact line, the boundaries are solids of revolution about an axis passing through the centre of the sphere. 
It is straightforward to apply this criterion to find the force-free equilibrium states of capillary bridges between flat and curved walls~\cite{concus2001bridges,kusumaatmaja2010bridges,brinkmann2004,baratian2015shape,deRuiter2015stability,farmer2015asymmetric}, 
but also those of droplets in contact with suspended solid particles, 
such as Pickering emulsions~\cite{guzowski2011capillary} 
and liquid marbles~\cite{aussillous2001liquid,mchale2015liquid}. 

The surface energy of such truncated-sphere solutions is invariant upon a rotation of the solid boundaries about the centre of the sphere [see Fig.~\ref{fig:Truncated}]. 
If one denotes $X_{\rm e}$ the equilibrium position of the centre of the sphere relative to a reference frame fixed to the solid boundaries ({\it e.g.}, their centre of mass), 
then such a rotation is equivalent to a displacement of $X_{\rm e}$ relative to that reference frame. 
Because the surface energy does not depend on the position of the capillary surface relative to the frame of reference of the boundaries ({\it i.e.}, $F_{\rm e}$ is not a function of $X_{\rm e}$), 
a reconfiguration of the boundaries can result in a net translation of the capillary surface without the system incurring any work.

\begin{figure*}
\includegraphics[width=\textwidth]{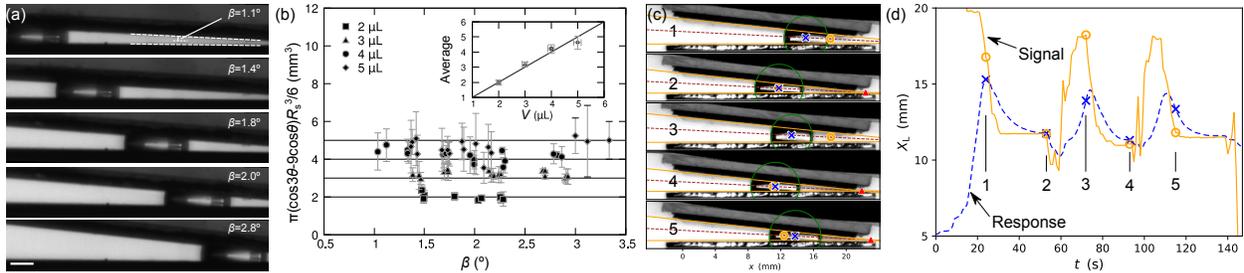}
\caption{\label{fig:WedgeShapes} (Color online) Shape invariance of droplets trapped in SLIPS wedges.
(a) A 4-$\mu{\rm L}$ water droplet equilibrates at different positions within a SLIPS wedge by adjusting the opening angle, $\beta$. 
(b) The radius of the droplet (normalised to eliminate variations in the apparent contact angle, $\theta = 100\pm 5^\circ$), is invariant upon changes in the 
opening angle of the wedge. The inset shows the dispensed and measured droplet volumes. 
Error bars correspond to the standard deviation of the sample. The scale bar is 1 mm.
(c) Manual actuation of a droplet by reconfiguration of the SLIPS geometry. The actuation signal shifts the position  the apex of the wedge (\textcolor{red}{$\blacktriangle$}). 
The new prescribed equilibrium position (\textcolor{YellowOrange}{$\circ$}) is followed by the centre  of the droplet (\textcolor{blue}{$\times$}). 
(d) Equilibrium position and droplet trajectory for the sequence shown in (c). The droplet trajectory, here tracked by measuring the position of the centre of the osculating sphere in the frame of reference of the lab, $X_{\rm L}$, follows the imposed signal with a lag determined by the friction force acting on the liquid. 
}
\end{figure*}

While this assertion is true in the quasi-static limit, 
more careful consideration is needed  to quantify the out-of-equilibrium contribution to energy dissipation arising from the motion of the boundaries. 
Consider the Lagrangian of the capillary surface in the over-damped limit, $L(X, t)=-U$, 
where $U=F(X)$ is the potential energy 
and $X$ is a coordinate describing its position relative to a set of solid boundaries. 
For a non-conservative system~\cite{galley2013classical}, the principle of minimisation of action leads to the classical Euler-Lagrange equation
\begin{equation}
- \frac{\partial L}{\partial X} + \nu \dot X = 0, 
\label{eq:EulerLagrange}
\end{equation}
where the second term on the left-hand side corresponds to the friction force and $\nu$ is the corresponding friction coefficient. 
Multiplying both sides of Eq.~(\ref{eq:EulerLagrange}) by $\dot X$ leads to an expression for the rate of change of the total energy, ${{\rm d} E}/{{\rm d} t } =  \partial  F/ \partial t - \nu {(\dot X)}^2.$
Close to equilibrium $F\approx F_{\rm e} + \frac{1}{2}k(X_{\rm e}) (X-X_{\rm e})^2$, 
with a spring constant $k$ that depends on the equilibrium configuration (here encoded through $X_{\rm e}$). 
Therefore, 
\begin{equation}
\frac{{\rm d} E}{{\rm d} t } =  \frac{\partial F_{\rm e}}{\partial t} +\frac{1}{2}\frac{\partial }{\partial t} \left[k(X-X_{\rm e})^2\right] - \nu {(\dot X)}^2.
\label{eq:EnergyExplicit}
\end{equation}
The first term in Eq.~(\ref{eq:EnergyExplicit}), $\partial F_{\rm e}/\partial t=0$,  confirms the energy invariance in quasi-static situations, where $X=X_{\rm e}$ and $\dot X=0$. 
The second and third terms give the contributions to energy dissipation due to small deviations from equilibrium and frictional forces, respectively. 

To quantify these contributions, we consider a slow sustained actuation of the boundaries over a timescale $\Delta t$ which results in a change in the equilibrium position, $\Delta X_{\rm e}$. 
Within our description this consists of prescribing an arbitrary function $X_{\rm e} (t)$ in an interval $0 \leqslant t \leqslant \Delta t$. 
Expressing the second term in the right-hand side of Eq.~(\ref{eq:EnergyExplicit}) using Eq.~(\ref{eq:EulerLagrange}), and integrating, gives the total energy consumption during the actuation,  
\begin{eqnarray}
 \Delta E  & = &  \int_{0}^{\Delta t} \frac{{\rm d} E }{{\rm d} t} {\rm d} t  \label{eq:EnergyChange} \\ 
 & = & \int_{0}^{1}  \Delta X_{\rm e} ^2 \left\{\frac{1}{2}\frac{\partial}{\partial T} \left[\left(\frac{\tau }{\Delta t}\right)^2 k {\dot x}^2\right] -  \left(\frac{\tau}{\Delta t}\right) k {\dot x}^2\right\} {\rm d}T,\nonumber
\end{eqnarray}
where we have defined the intrinsic relaxation timescale $\tau \equiv \nu/k$ and used the dimensionless variables $T\equiv t/\Delta t$ and $x\equiv X/\Delta X_{\rm e}$.  
The total energy consumption will vary depending on the actuation and response signals, $X_{\rm e}(t)$ and $X(t)$, subject to the initial condition $X(0)$.  
More importantly, both terms contributing to the energy change in Eq.~(\ref{eq:EnergyChange} )will be negligible whenever the actuation is slow relative to the relaxation timescale,
{\it i.e.,} if $\tau /\Delta t \ll 1$.

{\it Experiments.--}
Experimentally, a smooth transition between the energy-invariant states of a capillary surface upon boundary reconfiguration can only be achieved after eliminating contact-angle hysteresis~\cite{deRuiter2015stability}. 
Furthermore, to achieve the regime of negligible energy consumption during the reconfiguration~[see Eq.~(\ref{eq:EnergyChange})], one needs a test bed to determine the relaxation timescale of the 
out-of-equilibrium motion of the fluid. 
In our experiments, we used Slippery Liquid Infused Porous Surfaces (SLIPS)~\cite{wong2011bioinspired}, also known as lubricant-impregnated surfaces~\cite{smith2013droplet}, as a means of eliminating contact-angle hysteresis. 
When placed on a SLIPS surface, a small water droplet adopts a spherical shape with an apparent contact angle, $\theta=100^\circ \pm 5^\circ$.
We also observed a wetting ridge close to the intersection between the droplet and the SLIPS surface, indicating the presence of a lubricant layer coating the droplet and preventing 
direct contact between the droplet and the underlying solid~\cite{smith2013droplet,schellenberger2015direct,guan2015evaporation}.
This was further confirmed by measurements of the effective surface tension of a water droplet coated with a thin lubricant layer, $\gamma \approx 63\ {\rm mN\ m^{-1}}$,  
and of extremely low sliding angles ($<1^\circ$) for the droplet when tilting the SLIPS surface~[see Supplementary Information for more details]. 

To illustrate the formation of truncated-sphere droplet shapes in contact with SLIPS surfaces, we created a wedge of variable opening angle, $\beta$, by mounting two SLIPS on an adjustable stage. 
For such a configuration, the free energy of a droplet in contact with the boundaries, Eq.~(\ref{eq:F}), reduces to $F_{\rm e} =\gamma \pi (\cos3\theta-9\cos\theta)R_{\rm s}^2/3 + {\rm const.}$, 
with a sphere radius $R_{\rm s} = \{6V/[(\pi(\cos3\theta-9\cos\theta])\}^{1/3}$. 
The natural frame of reference of the solid boundaries is the apex of the wedge, from which the equilibrium position of the centre of the truncated spherical droplet along the bisector line is given by
$X_{\rm e} = - \cos \theta R_{\rm s}/\sin \beta$. 
In Fig.~\ref{fig:WedgeShapes}(a) we present equilibrium droplet configurations where $X_{\rm e}$ is varied by adjusting the angle of the wedge in the range $1.1^\circ \leqslant \beta \leqslant 2.8^\circ$.
Note that, because for small wedge angles $X_{\rm e}\sim 1/\beta$, the droplet can be displaced several times its own width along the bisector line upon a relatively small reconfiguration of the boundaries. 
In Fig.~\ref{fig:WedgeShapes}(b) we present measurements of the cube of the droplet radius, $R_{\rm s}$, as a function of the wedge angle. 
Because the apparent contact angle can vary from one set of SLIPS surfaces to another by a few degrees, we present our data absorbing the dependence on the contact angle, confirming the invariance of the truncated spherical shape upon changes in the orientation of the boundaries. 
This is equivalent to comparing the volume of the truncated sphere to the measured volume of the droplet, as shown by averaging the data over the wedge angle [see inset in Fig.~\ref{fig:WedgeShapes}(b)].  

In Fig.~\ref{fig:WedgeShapes}(c) we present a sequence of droplet configurations obtained by manually imposing an arbitrary signal $X_{\rm e}(t)$ [see also Supplementary Video S1]. 
In the absence of a threshold pinning force to overcome, a sudden change in the wedge geometry results in the immediate motion of the droplet towards a new equilibrium configuration. 
Therefore, the droplet's trajectory follows the imposed signal [Fig.~\ref{fig:WedgeShapes}(d)], with a lag determined by the interplay between the actuation and relaxation timescales.  

\begin{figure*}
\includegraphics[width=\textwidth]{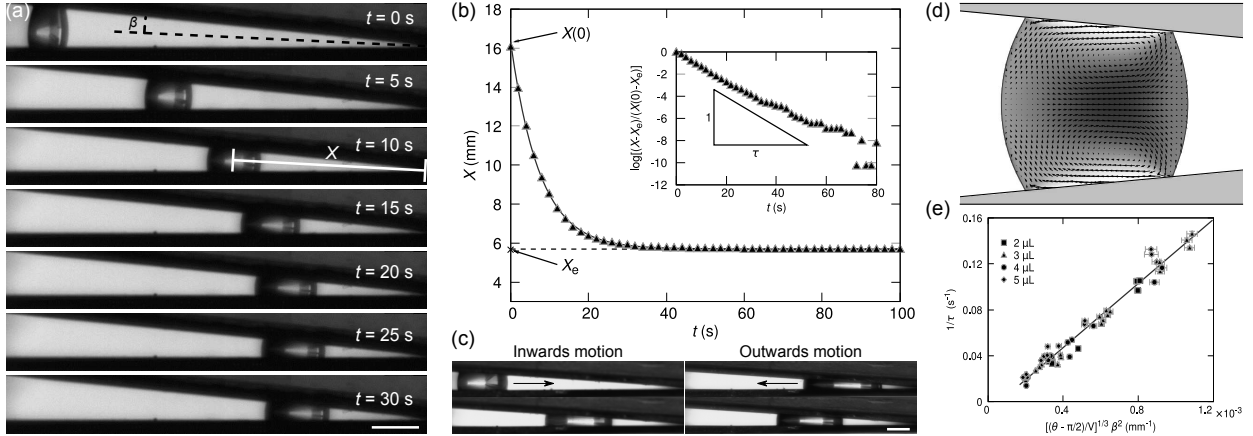}
\caption{\label{fig:WedgeRelaxation} Test bed for measuring the translational motion friction coefficient on SLIPS surfaces. (a) Time-lapse sequence of a 3-$\mu$L droplet moving inwards in a SLIPS wedge of opening angle $\beta = 2.8^\circ$. 
(b) Time dependence of the position of the droplet, tracked by measuring the average distance of the leading and trailing menisci relative to the apex of the wedge, $X(t)$. The time evolution obeys an exponential decay (continuous line) with a relaxation timescale $\tau$ (inset).
(c) Inwards and outwards equilibration of droplets of the same volume ($V=18~\mu$L) in wedges of the same angle, $\beta=2.0^\circ$. The direction of motion is indicated by the arrow. In equilibrium (bottom panels), both droplets select the same configuration. 
(d) Flow pattern of an inwards moving droplet (Lattice-Boltzmann simulation). The arrows indicate the velocity field in the frame of reference of the centre of the droplet. 
The grey scale indicates the local speed. 
(e) Scaling of the relaxation time of the droplet with droplet volume, apparent contact angle and wedge angle. The solid line is the expected scaling predicted by the theory (see text).
The scale bars are 2 mm.}
\end{figure*}

As a test bed to quantify the relaxation time of the translational motion of the droplet, $\tau$, we carried out experiments where one of the SLIPS surfaces is slowly brought into contact with a droplet. 
Upon contact, the droplet is allowed to relax to its equilibrium configuration [Fig.~\ref{fig:WedgeRelaxation}(a)].
In terms of our mathematical model, this corresponds to setting $X(0) = X_{\rm e} + \Delta X$, where $X_{\rm e}$ and $\Delta X$ are constants. 
Therefore, from Eq.~(\ref{eq:EulerLagrange}),  we expect an exponential relaxation, $X(t) = X_{\rm e} + \Delta X\exp(-t/\tau)$. 
Fig.~\ref{fig:WedgeRelaxation}(a) shows a typical experimental sequence of the relaxation process, where $X(0)>X_{\rm e}$.
The droplet moves inwards, and follows a remarkably smooth dynamics [Fig.~\ref{fig:WedgeRelaxation}(b)]. 
Fixing the initial position of the droplet within the wedge, to either $X(0)<X_{\rm e}$ or $X(0)>X_{\rm e}$, respectively leads to outwards and inwards motions, 
always resulting in the same stable equilibrium state [Fig.~\ref{fig:WedgeRelaxation}(c)]. 

Because the SLIPS surfaces eliminate contact line friction~\cite{wong2011bioinspired,smith2013droplet}, 
the friction force, $-\nu \dot X$, results from the flow within the droplet, the lubricant layer, and the wetting ridge. 
{This situation is similar to the case of a wetting capillary bridge moving within solid wedge~\cite{reyssat2014drops}.}
The contribution of the lubricant layer, relative to the bulk of the droplet, scales as $h/H\approx10^{-2}$, where $h \approx 10\ \mu {\rm m}$ is the thickness 
of the lubricant layer~\cite{guan2015evaporation} and $H\approx 1\ {\rm mm}$ is the typical thickness of the droplet, and is therefore negligible. 
We expect that the contribution from the ridge is also negligible, as the apparent angle is always close to its static value during the relaxation of the droplet. 
Therefore, we assume that the dominant contribution to the friction force comes from the flow pattern within the droplet. 
To gain insight into the structure of the flow, we carried out Lattice-Boltzmann simulations~\cite{desplat2001ludwig} of 2D droplets equilibrating in wedge geometries~[see Supplementary Information for details]. 
The simulations reveal a pressure-driven flow within the droplet, similar to a Jeffery-Hamel flow~\cite{rosenhead1939steady}, truncated at the leading and trailing menisci, which move 
at uniform speed [Fig.~{\ref{fig:WedgeRelaxation}}(d)].
This effect can be captured by considering a slip length, $\ell$, which quantifies the lubrication imparted by the SLIPS surface on the motion of the apparent contact lines.   
After some manipulations, the expected friction coefficient can be expressed as $\nu \approx 12\mu V/(1+6\epsilon)H^2$, which is the familiar result for a Poiseuille flow 
with a correction that depends on the slip effect, where $\epsilon = \ell/H$~[see Supplementary Information].

To compare the theoretical prediction to the measured relaxation times, we use a model for the out-of-equilibrium droplet morphology 
assuming a quasi-spherical barrel shape intersecting the solid at the apparent contact angle, $\theta$~\cite{Gutierrez2017}. 
The droplet shape can then be used to construct the energy landscape, $F(X)$, which in turn fixes the spring constant, $k$. 
To leading order in $\beta$ and $\theta-\pi/2$ (corresponding to the regime of our experiments), we obtain $k \approx 3\pi\gamma\beta^2/(\theta-\pi/2)$~[see Supplementary Information].  
Using the geometrical relation $H \approx (4V/\pi)^{1/3}(\theta-\pi/2)^{2/3}$, we  then find a prediction for the relaxation time $\tau=\nu / k \approx [\mu/\gamma(1+6\epsilon)\beta^2][4V/\pi(\theta-\pi/2)]^{1/3}$. 
Fig.~\ref{fig:WedgeRelaxation}(e) confirms the scaling of $\tau$ with $V$, $\theta$ and $\beta$. 
Using the measured surface tension of the lubricant-cloaked droplets, $\gamma =63\ {\rm mN\ m^{-1}}$, and the reported value of the viscosity of water at room temperature, $\mu=1\  {\rm mPa\ s}$, 
the only unknown parameter in our prediction is the slip-length to drop-height ratio, $\epsilon$.
We find a best fit to the data by choosing $\epsilon \approx 0.2$, which corresponds to a $\approx 60\%$ drag reduction relative to the reference Jeffery-Hamel flow. 

\begin{figure}[t!]
\includegraphics[width=0.5\columnwidth]{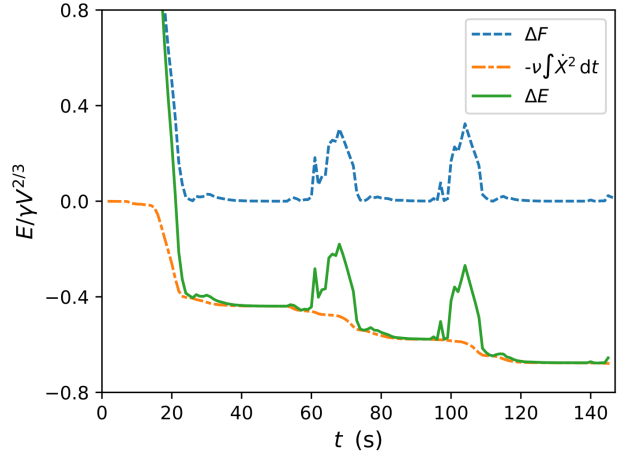}
\caption{\label{fig:EnergyChange}(Color online) Change in energy of a droplet upon manipulation within a SLIPS wedge (solid line). A fast actuation (intermediate peaks) leads to changes in the potential energy, $\Delta F$ (dashed line); a slow actuation (intermediate plateaus) gives rise to small changes in the energy due to dissipation upon equilibration, $- \nu \int {\rm d} t {\dot X}^2$ (dashdotted line).}
\end{figure}

Our measurements of the relaxation time allow us to calculate the friction coefficient, and therefore, to estimate the typical friction force experienced by the droplets upon actuation. 
For the actuation sequence shown in~Fig.~\ref{fig:WedgeShapes}(d), $\nu \approx 0.013~{\rm mN~s~mm^{-1}}$, and the droplets move with velocities ranging from $ -0.4~{\rm mm~s^{-1}}$ to $1.2~{\rm mm~s^{-1}}$.
Therefore, the friction force varies within $-0.005~{\rm m N}$ -- $0.015~{\rm mN}$, and is significantly smaller than the weight of a droplet of equal size ($\approx 0.04~{\rm m N}$). 
This implies that a relatively weak driving is enough to achieve translational motion, even out of equilibrium. 
This is evidenced in Fig.~\ref{fig:EnergyChange}, where we present the corresponding reconstructed energy change, Eq.~(\ref{eq:EnergyChange}). 
After the droplet has equilibrated, the total energy, $E$, is always reduced due to the dissipation term, $-\nu \int {\rm d} t {\dot X}^2$. 
However, the energy only changes significantly when the actuation timescale is much faster than the relaxation timescale of the drop. 
These `fast' events appear as intermediate peaks in the potential energy, where the system is driven out of equilibrium, and correspond to the segments 2--3 and 4--5 in Fig.~\ref{fig:WedgeShapes}(d).  
During the rest of the actuation, where the driving is relatively slow, the change in potential energy and the energy dissipation remain negligible [plateaus in Fig.~\ref{fig:EnergyChange}],  
confirming that it is possible to approach the limit of an energy-invariant translation of the droplet upon a slow reconfiguration of the boundaries.

Our results thus open up the possibility of developing pathways for droplet actuation at no potential energy cost and involving low energy dissipation. 
We highlight the relevance of these ideas in the future development of contact-free microfluidic channels that overcome both contact-line pinning and reduce viscous friction using liquid-layer mediated slip. 
These principles can be extended to treat multiphase systems such as encapsulated droplets, solid particles and even cells, and can have a wider relevance 
in tribology~\cite{urbakh2004nonlinear,he1999adsorbed} to encourage the development of technologies that remove the minimum force necessary to create motion and achieve the accurate manipulation of target objects.

Here we have focused on capillary surfaces of spherical symmetry and in contact with solids of uniform wettability as a means to illustrate energy invariance upon boundary reconfiguration.
These ideas, however, can also be applied to study capillary surfaces of a different symmetry and in contact with boundaries of prescribed wettability distributions, 
opening the possibility of designing target energy landscapes for liquids in contact with solids as the basis for new kinds of `capillary metamaterials'. 

We thank F. Mugele and C. Semprebon for useful discussions. 
JHG thanks REECE Innovation for financial support and D. Wood and J. Martin for useful discussions. 
JHG and ER-G acknowledge support from Northumbria University {\it via} PhD studentships.

%


\end{document}